\documentclass[IEEEtran,twocolumn]{revtex4}
\usepackage{graphicx,epsfig}
\usepackage{amsmath,amssymb,revsymb}
\usepackage{latexsym}

\def\duzomniejsze{<\kern-.7mm<}
\def\duzowieksze{>\kern-.7mm>}

\def\textbf#1{{\bf #1}}
\def\beq{\begin{equation}}
\def\eeq{\end{equation}}

\def\be{\begin{equation}}
\def\ee{\end{equation}}
\def\eea{\end{eqnarray}}
\def\bea{\begin{eqnarray}}

\def\ben{\begin{eqnarray}}
\def\een{\end{eqnarray}}
\def\beqa{\begin{eqnarray}}
\def\eeqa{\end{eqnarray}}

\newcommand{\bei}{\begin{itemize}}
\newcommand{\eei}{\end{itemize}}
\newcommand{\bee}{\begin{enumerate}}
\newcommand{\eee}{\end{enumerate}}

\def\tr{{\rm Tr}}

\def\>{\rangle}
\def\<{\langle}

\def\e{\epsilon}
\def\ra{\rightarrow}
\def\la{\leftarrow}

\def\ot{\otimes}

\def\eass{E_{\rm a}}

\newcommand{\msec}[1]{Sec.~\ref{sec:#1}}
\newcommand{\fig}[1]{Fig.~\ref{fig:#1}}
\newcommand{\eq}[1]{Eq.~(\ref{eq:#1})}

\newtheorem{definition}{Definition}

\newlength{\blank}
\begin{document}

\title{Quantum network communication -- the butterfly and beyond} 

\begin{abstract}

We study the problem of $k$-pair communication (or multiple unicast
problem) of quantum information in networks of quantum channels.
We consider the asymptotic rates of high fidelity quantum
communication between specific sender-receiver pairs.
Four scenarios of classical communication assistance (none, forward,
backward, and two-way) are considered.
{\sc(i)} We obtain outer and inner bounds of the achievable rate
regions in the most general directed networks.
{\sc(ii)} For two particular networks (including the butterfly
network) routing is proved optimal, and the free assisting classical
communication can at best be used to modify the directions of quantum
channels in the network.
Consequently, the achievable rate regions are given by counting edge
avoiding paths, and precise achievable rate regions in all four
assisting scenarios can be obtained.
{\sc(iii)} Optimality of routing can also be proved in classes of
networks.  The first class consists of directed unassisted networks in
which (1) the receivers are information sinks, (2) the maximum
distance from senders to receivers is small, and (3) a certain type of
$4$-cycles are absent, but without further constraints (such as on the
number of communicating and intermediate parties).  The second class
consists of arbitrary backward-assisted networks with $2$
sender-receiver pairs.
{\sc(iv)} Beyond the $k$-pair communication problem, observations are
made on quantum multicasting and a static version of network
communication related to the entanglement of assistance.

\end{abstract}
\author{Debbie Leung$^{(1)}$, Jonathan Oppenheim$^{(2)}$ and Andreas
Winter$^{(3)}$}
\affiliation{$^{(1)}$Institute for Quantum Computing, University of
Waterloo, Waterloo, Canada}
\affiliation{$^{(2)}$Department of Applied Mathematics and
Theoretical Physics, University of Cambridge, Cambridge, U.K.}
\affiliation{$^{(3)}$Department of Mathematics, University of Bristol,
Bristol, U.K.}
\maketitle

\parskip=1ex
\parindent=0ex

\section{Introduction}

Consider a network of point-to-point communication channels.  
At any time period, we model the set of users sharing access to the 
same point as a party.  
Most generally, any party may want to transmit data to a set of other
parties, and such data can be correlated in time and space.
A central question is whether a given network can handle a specific
joint communication task.
For example, in the {\em multicast} problem, one party wants to send
the same data to a specific list of other parties (the most common
example is sending invitations to a wedding).
In contrast, the {\em multiple unicast} or the $k$-pairs communication
problem is concerned with $k$ specific (disjoint) sender-receiver
pairs, who are trying to communicate $k$ {\em independent} messages in
the given network.

We consider networks of parties connected by noiseless channels, and
they are represented by vertices and edges in a graph.  Each edge is
weighted by the capacity of the corresponding channel, and its
direction (if any) follows that of the channel.  

Network communication was traditionally done by ``routing'' (also
known as the store-and-forward method) in which received data is
simply copied and forwarded without data processing.
In 2000, Ahlswede, Cai, Li, and Yeung \cite{ACLY00} provided the first
example that nontrivial coding of data can strictly improve the
communication rate for multicasting in the directed ``butterfly
network'' (see \msec{butterfly}).
The coding method in \cite{ACLY00} also applies to the $2$-pair
communication problem (formalized in \cite{YZ99}) in the same network
(see the discussions in \cite{LL04a,HKL04}) demonstrating the general
advantage of network coding for the $k$-pair communication problem in
directed networks.
For undirected networks, it was conjectured that routing is optimal
for the $k$-pair communication problem \cite{LL04b,HKL04} and it was 
proved in many cases, such as when $k \leq 2$ \cite{LL04a}, and others
\cite{JVYY06,HKL06}.

This paper is concerned with quantum communication through quantum
networks.  We primarily focus on the $k$-pair communication problem.
Our goal is to find the optimal achievable rates (given by the
boundary of a $k$-dimensional achievable rate region).


In our study of high fidelity quantum communication through an
asymptotically large number of uses of the (quantum) butterfly
network, routing turns out optimal.  This contrasts with the advantage
of network coding in the classical setting, and demonstrates 
another difference between quantum and classical information.
Thus, quantum information flowing through this communication network
resembles a classical commodity more than classical information.  We
believe that such behavior holds for general networks, and provide
reasons why it is true for a certain class of ``shallow'' quantum
networks in which the maximum distance between any sender-receiver
pair is small.
We also study communication scenarios with various auxiliary resources
and optimality of routing is essentially unchanged.
%
In particular, free classical {\em back} communication effectively
makes the quantum channel undirected, and our optimality proof of
routing provides some partial answer to the question raised in
\cite{LL04b} in the quantum setting.

Our work was inspired by the earlier, complementary, study of Hayashi,
Iwama, Nishimura, Raymond, and Yamashita on the quantum butterfly
network \cite{HINRY06}.  They fix the quantum communication rates as
in the classical case, and optimize the fidelity of the transmitted
states.  Deviation from the classical case is manifest in that the
optimal $1$-shot fidelity is upper bounded by $0.983$.  
During the preparation of this manuscript, we found that Shi and
Soljanin have studied a quantum version of multicasting in quantum
network \cite{SS06} that is complementary to our study.
After the initial submission of this manuscript to the eprint server
\cite{LOW06}, Hayashi studied the case of $2$-pair communication
problem in the directed butterfly network with entanglement shared
between the senders, a setting that is also complementary to the 
current one.

We shall begin in Section \ref{sec:butterfly} with the butterfly
network as a motivating example.  
%
%
Starting from this simpler case, we formalize the network
communication problem of interest and review useful techniques, and
discuss their generalizations.
Then, we focus back on the butterfly network, summarize the classical
solution in \msec{cbutterfly} and present our optimal quantum
communication protocols for scenarios with differing free auxiliary
resources in \msec{qbutterfly}.  Another example is discussed in
\msec{examples} which will further demonstrate our results for more
general networks presented in \msec{general}: an optimality proof for
routing of quantum information in certain shallow networks
(\msec{shallow}), outer and inner bounds of the achievable rate region
for the $k$-pair communication problem in the most general network
(\msec{mincut}), an optimal solution for the $2$-pair case assisted by
back classical communication (\msec{ghost}), and a reduction of the
entanglement assisted case to the classical information flow problem
(\msec{eassist}).  We discuss two other quantum network communication
problems in \msec{others}: (1) a quantum analogue of the multicasting
problem -- sharing a cat-state between a reference and $k$ receivers
-- and (2) network communication based on a ``static'' quantum
resource -- a pure quantum state shared by the parties -- assisted by
$2$-way classical communication.  We conclude with some open problems
in \msec{conclusion}.

We use the following notations throughout the paper.  The resource of
being able to send a classical bit noiselessly from one party to
another is called a {\em cbit}.  A state in a $2$-dimensional Hilbert
space is called a {\em qubit}, and the ability to transmit it is
called a {\em qbit}.
The quantum analogue of a shared random bit is called an {\em ebit} --
the resource of two parties sharing a copy of the joint state ${1
\over \sqrt{2}} (|00\>+|11\>)$.  An ebit can be created using 
other resources (say, qbits, or other quantum states) and be consumed
to generate other resources.  For example, in teleportation, $2$ cbits
and $1$ ebit generate $1$ qbit \cite{BBCJPW93}, and in superdense
coding $1$ ebit and $1$ qbit generate $2$ cbits$\;$\cite{BW92}.  

\section{Motivating example -- the butterfly network}
\label{sec:butterfly}

\begin{figure}[h]
\includegraphics[width=3in]{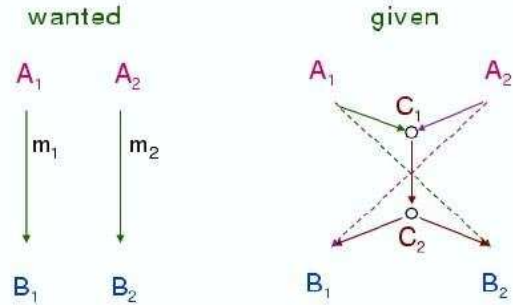}
 \caption{The butterfly network.  The left diagram represents the task 
 to be achieved, the right diagram represents the available resources.}
\label{fig:bf-setup}
\end{figure}

{\bf Setting for butterfly network:} Consider two senders $A_1$ and
$A_2$, who want to send two independent messages $m_1$ and $m_2$ to
two respective receivers $B_1$ and $B_2$.  Available to them is a
network of $7$ {\em noiseless} directed channels and two helpers $C_1$
and $C_2$ depicted in Fig.\ \ref{fig:bf-setup}.
For each call to the network, each channel in the network can be used
once.  The number of calls to the network represents our ``cost'' to
be minimized.  (The network is charged as a package.)  Local resources
are free.
In the classical (quantum) setting, both messages and the available
channels are classical (quantum).

\begin{definition}[Rate region for butterfly network]
In the asymptotic scenario, we allow large number of calls to the
network.  Let ${\cal P}_n$ denote a protocol that uses the network $n$
times along with other allowed resources, and communicates $m_1,m_2$
of sizes $n (r_1{-}\delta_n),n (r_2{-}\delta_n)$ bits/qubits with
fidelities at least $1{-}\epsilon_n$ for $\delta_n, \e_n \ra 0$.
Then, we say that the rate pair $(r_1,r_2)$ is achievable.  The
achievable rate region is the set of all achievable rate pairs.
\end{definition} 

{\bf General setting:} We consider communication networks in which the
number of sender-receiver-pairs and intermediate parties and the
capacities of the channels connecting them are arbitrary.  To be
concrete, consider $k$ senders $A_1$, $\cdots$, $A_k$, who want to
send $k$ independent messages $m_1$, $\cdots$, $m_k$ to $k$ respective
receivers $B_1$, $\cdots$, $B_k$.  Available to them is an arbitrary
directed network of noiseless channels and intermediate helpers $C_1$.
The rest of the setting is the same as that in the butterfly network
and the achievable rate region for such ``$k$-pair communication''
problem is defined analoguously.

\begin{definition}[Rate region for general network]
In the asymptotic scenario, we allow large number of calls to the
network.  Let ${\cal P}_n$ denote a protocol that uses the network $n$
times along with other allowed resources, and communicates
$m_1,\cdots,m_k$ of sizes $n (r_1{-}\delta_n),\cdots,n (r_k{-}\delta_n)$
bits/qubits with fidelities at least $1{-}\epsilon_n$ for $\delta_n,
\e_n \ra 0$.  Then, we say that the rate $k$-tuple $(r_1,\cdots,r_k)$ is
achievable.  The achievable rate region is the set of all achievable
rate $k$-tuples.
\end{definition} 

We now discuss aspects of the general problem.  

Note that in the asymptotic setting, imposing time ordering of the
usage of the channels does not affect the achievable rate region.  

We choose a measure of fidelity that achieves the strongest notion of
approximation.  
We are concerned with sending quantum messages through networks of
quantum channels.  In this setting, we require the protocol to
transmit a message in a way that preserves arbitrary entanglement
between it and any reference system.  In other words, the joint state
held by the receiver and the reference after the protocol should be
close in trace distance to that held by the sender and the reference
before the protocol.  In the specific cases solved in this paper, the
achieving optimal protocols turns out to be exact.  In our proofs of
optimality (obtaining outerbounds of the rate region), we give full 
consideration of protocols that have small errors.  

We collect tools and techniques that are useful for the general 
$k$-pair communication problem, and
occasionally refer to Fig.\ \ref{fig:bf-setup} as an example.

\begin{enumerate} 

\item{Exact rate regions and optimal protocols via matching inner and
outer bounds}
\label{match}

Throughout the paper, whenever possible, we (1) describe simple
protocols and the corresponding inner bounds for the rate region and
(2) obtain outer bounds that match the inner bounds.  Each outer bound
has to be completely general, and applies asymptotically.  Altogether,
these two steps give the exact achievable rate region and prove the
optimality of the simple protocols described.

\item{Convexity and monotonicity of achievable rate regions} 

Note that if a rate pair $(r_1,r_2)$ is achievable, so is any
$(r_1',r_2')$ with $r_1' \leq r_1$ and $r_2' \leq r_2$.  Also, the
convex hull of a set of achievable rate pairs are also achievable by
time sharing of the underlying protocols.  Similarly for the $k$-pair
communication problem.

\item{Outer bounds by cuts}
\label{min} 

Consider a bipartite cut, i.e., a partition of the vertices into two
disjoint subsets of parties $S_1$ and $S_2$.  
We can bound the sum of communication rates from all parties in $S_1$
to all parties in $S_2$ by adding the capacities of all forward
communication channels from $S_1$ to $S_2$ (since grouping the parties
together can only increase the communicate rate and back communication
does not help \cite{CDNT97}).
This induces a bound on the sum of rates for the pairs each with the
sender in $S_1$ and the receiver in $S_2$.
  
For example, let $S_1 = \{A_1, B_2, C_1\}$ and $S_2 = \{A_2, B_1,
C_2\}$ in Fig.\ \ref{fig:bf-setup}.  Then, we can bound $r_1$, because
any protocol on the butterfly network communicating from $A_1$ to
$B_1$ will also communicate at least the same amount of data from
$S_1$ to $S_2$.
There is only $1$ forward channel from $S_1$ to $S_2$, so $r_1 \leq 1$.

We will also see scenarios in which the channels are effectively
undirected.  In those cases, the total communication rate from all the
parties in $S_1$ to those in $S_2$ is upper bounded by the total
capacities of all the channels between them.

\item{Inner bounds via the max-flow-min-cut theorem} 

By the max-flow-min-cut theorem \cite{FF56}
edges crossing a min-cut can be extended to edge-avoiding paths 
leading from a sender to a receiver. 

\item{Sizes of significant shares and quantum parts in
quantum-classical dual compression}
\label{qssqc} 

We will make use of two lower bounds for the sizes of the individual
communicated parts when quantum data is sent in a distributed manner.

(a) A quantum secret sharing scheme is an encoding of a quantum state
(the secret) in a multiparty system.  Each party owns one system
called a ``share.''  Authorized sets of parties can reconstruct the
secret (with high fidelity), while unauthorized sets of parties can
learn negligible information about the secret.  A share $S_s$ is
``significant'' if there exists an unauthorized set $S_u$ such that
$\{ S_s \} \bigcup S_u$ is authorized.  It was proved in
\cite{Gottesman99qss} that for exact schemes (reconstruction and
hiding are perfect), the size of any significant share is at least the
size of the quantum secret.  An alternative proof of this result in
\cite{IMNTW03} extends to the near-exact case.  

More precisely, let $S(\cdot)$ denote the von Neumann entropy, and
$I^{\rm coh}(S_1 \> S_2) = S(S_2) - S(S_1S_2)$ denote the coherent
information from $S_1$ to $S_2$.  Let $S$ be the secret, purified by
the reference system $R$.  Then,
\begin{equation} 
 S(S_s) \geq  S(S) - (\epsilon' + \gamma)/2
\end{equation} 
where $\epsilon'$ and $\gamma$ are the respective upper bounds on
$I^{\rm coh}(R \> S) - I^{\rm coh}(R \> \tilde{S})$ and the quantum
mutual information $I(S_u{:}R)$, and they are both negligible when
recovery of the secret is near-exact.  For completeness, the proof in
\cite{IMNTW03} is duplicated in the endnote \cite{supp1}, with
$\epsilon'$ and $\gamma$ explicitly derived and inserted.

(b) A quantum-classical dual compression scheme encodes a quantum
source into a quantum part and a classical part.  It was proved in
\cite{BHJW00} that the quantum part cannot be smaller than the von
Neuman entropy of the source.


(c) We will also use an immediate consequence of Theorem $6$ in
\cite{Gottesman99qss} that logical transformation of the encoded
quantum secret can be performed by operating on an authorized set
without involving other shares.  We prove an extension of this result
in the endnote \cite{supp2} and give a precise statement here.

Let $S$ be the system holding the secret and $R$ be its purifying
reference system.  Let $W$ be an isometry encoding $S$ into systems
$A$, $B$, and $D$ where $D$ is discarded (to allow the possibility of
mixed state secret sharing schemes).  Suppose the encoding is
invertible on $A$ with error $\epsilon$ (i.e., $\exists Y$ an isometry
taking $A$ to $\tilde{S} E$ such that $R\tilde{S}$ is in a state
$\epsilon$-close in trace distance to what's originally in $RS$).
Then, a desired operation $U$ to be applied to the secret $S$ (before
the encoding) can be performed with error $2 \epsilon$ by
applying $Y^\dagger (I_E \otimes U_{\tilde{S}}) Y$ to $A$ alone.

We state this result for unitary $U$ but our proof in the endnote
\cite{supp2} holds if we replace the unitary $U$ by an arbitrary
quantum operation on $S$ and $\tilde{S}$.  Also, compared to
\cite{Gottesman99qss}, the current proof is constructive and
operational -- it simply asserts that the intuitive approach of
``decoding the secret, keeping the auxiliary system $E$, operating on
the decoded state, and reversing the decoding'' works in a way that
preserves the correlation with the remaining shares.

\end{enumerate}

We will now use this set of general techniques to investigate our 
example, the butterfly network.

\subsection{Classical case \cite{ACLY00}}  

\label{sec:cbutterfly}

In the classical case, one use of each channel in the network
communicates $1$ classical bit, and the messages $m_1, m_2$ are
classical bit strings.  

\begin{figure}[h]
\includegraphics[width=2in]{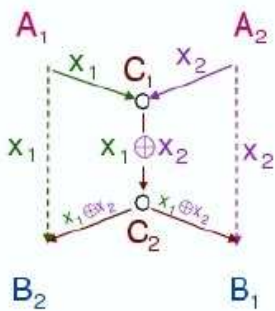}
 \caption{The optimal protocol for the classical butterfly network.
 In the above, and from now on, we rearrange the locations of 
 $B_1,B_2$ to improve diagrammatic clarity.}
\label{fig:bf-c-sol}
\end{figure}

{\bf Inner bound:}
Let $x_{i}$ be the $1$-bit message to be communicated from $A_{i}$ to
$B_i$ for $i=1,2$.
A method that simultaneously communicates $x_{1,2}$ with exactly $1$
call to the network is given by Fig.\ \ref{fig:bf-c-sol}.

{\bf Outer bound:} 
The above protocol turns out to be optimal because we can prove
matching outer bounds $r_1 \leq 1$ and $r_2 \leq 1$, using the min-cut
method.  
To show $r_1 \leq 1$, consider the bipartite cuts $S_1 = \{A_1, B_2,
C_1\}$ and $S_2 = \{A_2, B_1, C_2\}$.  The bound follows from the fact
that there is only one forward channel from $S_1$ to $S_2$.  (See
the detail argument in item \ref{min} in the previous subsection.)
A similar argument with the cut $S_1 = \{A_2, B_1, C_1\}$ and $S_2 =
\{A_1, B_2, C_2\}$ shows $r_2 \leq 1$.

Since the outer and inner bounds are matching, by item \ref{match},
the $1$-shot, exact, protocol in Fig.\ \ref{fig:bf-c-sol} is indeed
optimal, and the rate region is just the unit square.  (See Fig.\
\ref{fig:bf-sum}.)

This example illustrates some common features in network communication
-- a ``bottleneck'' from $C_1$ to $C_2$ and channels that go to the
``wrong places.''  It also exhibits how nontrivial coding techniques
can be applied to improve the communication rates for ``information
flow'' in networks, beyond simple routing.

\subsection{Quantum case} 
\label{sec:qbutterfly}

The setting is the same as the classical case, except now the messages
$m_1, m_2$ are uncorrelated quantum states $|\psi_1\>, |\psi_2\>$, and
each use of the channel allows the communication of $1$ qubit.  
(To simplify notations, we denote inputs as pure states, but since
 communication is entanglement-preserving, the discussion applies to
 sending parts of entangled states by linearity.)

Clearly the classical coding strategy depicted in Fig.\
\ref{fig:bf-c-sol} fails in the quantum case -- the encoding by
$A_1,A_2,C_2$ involves cloning unknown quantum states, and quantum
analogues of the $\oplus$ operation do not provide the desired result.
In fact, \cite{HINRY06} showed that if one demands one qubit states
$|\psi_1\>, |\psi_2\>$ to be communicated by one use of the network,
the fidelity is upper bounded by 
$0.983$ (though better than $0.52$).

In the following, we will consider an asymptotic number of calls of
the network, and demand high fidelity transmission, and optimize the
achievable rates.  We consider five different scenarios of free
auxiliary resources (also known as assisting resources).  We first
consider the no assistance case, followed by the easier case of having
free backward classical communication (which turns out to be no worse
than free two-way classical communication).  Then, we consider the
more intricate case of having free forward classical communication,
and finish off with the entanglement assisted case.

\vspace*{1ex} 

$\bullet$ {\bf Unassisted case (no free resource)} 

\vspace*{1ex} 

{\bf Inner bound:} The rate pair $(r_1,r_2) = (1,0)$ is achieved by
sending $|\psi_1\>$ from $A_1$ to $C_1$ to $C_2$ and finally to $B_1$.
The rate pair $(r_1,r_2) = (0,1)$ is achieved by a similar protocol.
By time sharing and monotonicity, any point in the first quadrant with
$r_1 + r_2 \leq 1$ can be achieved.

{\bf Outer bound:} We will prove $r_1+r_2 \leq 1$.  

\begin{figure}[h]
\includegraphics[width=2.4in]{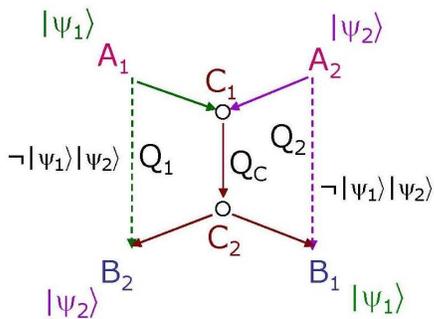}
 \caption{Proof ideas for the outer bound for the achievable rate
 region of the unassisted butterfly network.  ``$\neg
 |\psi_1\>|\psi_2\>$'' labels a state that is nearly independent of
 $|\psi_1\>|\psi_2\>$.}
\label{fig:bf-q-unassist}
\end{figure}

The main idea is captured in Fig.\ \ref{fig:bf-q-unassist}.  

Consider using the network $n$ times to enable $A_i$ to send a state
$|\psi_i\>$ of size $n_i = n (r_i-\delta_n)$ qubits for $i=1,2$ and for
$\delta_n \ra 0$.
Let $Q_1$, $Q_2$, and $Q_C$ be the quantum states sent using the $n$
channel-uses from $A_1$ to $B_2$, $A_2$ to $B_1$, and $C_1$ to
$C_2$ respectively.
We can consider $|\psi_1\>$ {\em and} $|\psi_2\>$ together as the
quantum secret, and apply item \ref{qssqc}(a) at the beginning of
\msec{butterfly}.
Clearly $\{Q_1,Q_2,Q_C\}$ is an authorized set.  We will now prove
that $\{Q_1,Q_2\}$ is unauthorized.  The basic idea is that, $Q_1$ has
to be independent of $|\psi_1\>$ if $B_1$ is to receive it faithfully.
It is also independent of $|\psi_2\>$ by causality.  Thus $Q_1$ is
independent of both $|\psi_{1,2}\>$.  

To capture this formally, the $A_1 \ra B_1$ message $M_1$ should be
described as half of a maximally entangled state with a reference
system, say, $R_1$.  The $A_2 \ra B_2$ message $M_2$ likewise has a
reference system $R_2$.  Let $R=R_1R_2$.
Let $S(\cdot)$ and $I(\cdot\,{:}\,\cdot)$ denote the von Neumann
entropy of a system and the quantum mutual information between two
systems (with the underlying state implicit).
By independence of the messages and causuality, $I(R_1\,{:}\,R_2) =
I(Q_1\,{:}\,Q_2) = I(Q_1 R_1\,{:}\,Q_2 R_2) = 0$.
Our observation above further says that $I(Q_1\,{:}\,R)$ and
$I(Q_2\,{:}\,R)$ are both small.
Now, 
\bea
& & \hspace*{-5ex}I(Q_1Q_2\,{:}\,R) 
\nonumber
\\
\nonumber
& {:}{=} & S(Q_1Q_2) + S(R) - S(Q_1Q_2R) 
\\
\nonumber
& = & S(Q_1){+} S(Q_2) + S(R_1) {+} S(R_2) - S(Q_1 R_1) {-} S(Q_2 R_2) 
\\
\nonumber
& {:}{=} & 
I(Q_1\,{:}\,R_1) + I(Q_2\,{:}\,R_2)
\eea 
which is small (the equality is due to the various independence
conditions).
Thus, $Q_C$ is a significant share, and it has at most $n$ qubits, and
applying item \ref{qssqc}(a), $n \geq n_1 + n_2$ 
which gives the desired bound.

With the matching inner and outer bounds, we conclude that the
achievable rate region is the triangle with vertices
$(0,0),(0,1),(1,0)$, and time sharing between the communication paths
$A_1 \ra C_1 \ra C_2 \ra B_1$ and $A_2 \ra C_1 \ra C_2 \ra B_2$ gives
the optimal protocol.  Note that the optimal communication protocol is
exact and entanglement preserving.

\vspace*{1ex} 

$\bullet$ {\bf Back-assisted case (with free backward classical 
communication)}

\vspace*{1ex} 

First, note that $2$ bits of back classical communication can be used
to reverse the direction of a qubit quantum channel: use the quantum
channel to create $1$ ebit, followed by teleportation in the reverse
direction.  Thus, free back communication makes quantum networks
undirected.  With this observation, we describe new communication
protocols for the butterfly network.

\begin{figure}[h]
\includegraphics[width=2in]{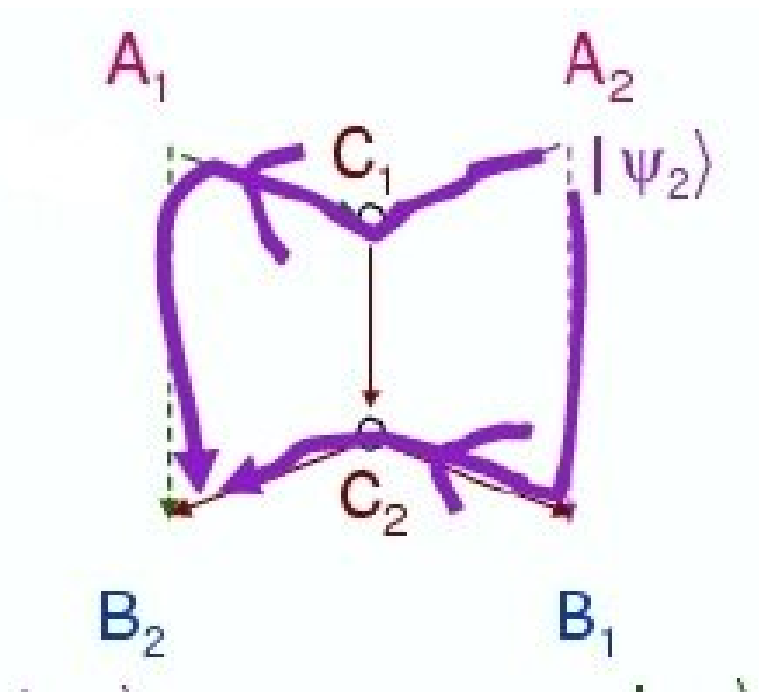}
 \caption{ The achieving protocol for the rate pair $(0,2)$ 
 for the back-assisted butterfly network.}
\label{fig:bf-q-b-ib}
\end{figure}

{\bf Inner bound:} The rate pair $(r_1,r_2) = (0,2)$ is achieved by an
exact, $1$-shot, protocol. $A_2$ sends one qubit along the path $A_2
\ra C_1 \ra A_1 \ra B_2$ and another qubit along the path $A_2 \ra B_1
\ra C_2 \ra B_2$ (see Fig.\ \ref{fig:bf-q-b-ib}).  These are
edge-avoiding paths, and thus, two qubits can be transmitted in a
single network-call.
Likewise, $(r_1,r_2) = (2,0)$ is also achievable, and so is the entire 
triangle with vertices $(0,0),(2,0),(0,2)$.  

{\bf Outer bound:} We will prove that $r_1+r_2 \leq 2$.  Consider the
cut $S_1 = \{A_1,B_2\}$ and $S_2=\{C_1,C_2,A_2,B_1\}$.  Let $r$ be the
maximum amount of entanglement between $S_{1,2}$ created per network
call.
There are only two channels across this cut, so, $r \leq 2$.
Now, any asymptotic $n$-use protocol on the butterfly network
communicating $n(r_i-\delta_n)$ qubits from $A_i$ to $B_i$ enables
$A_1,B_2$ and $A_2,B_1$ to share at least $n(r_1+r_2-2\delta_n)$ ebits
(with high fidelity).  Per network use, $r_1 + r_2 - 2\delta_n$ ebits
are created.
Altogether, taking large $n$ limit, $r_1 + r_2 \leq r \leq 2$.
\begin{figure}[h]
\includegraphics[width=2in]{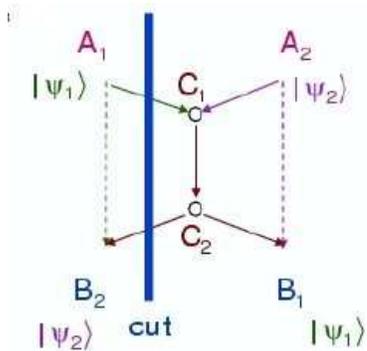}
 \caption{The proof idea for the outer bound of the achievable rate
 region of the back-assisted butterfly network.} 
\label{fig:bf-q-b-ob}
\end{figure}

\begin{figure}[h]
\includegraphics[width=3.3in]{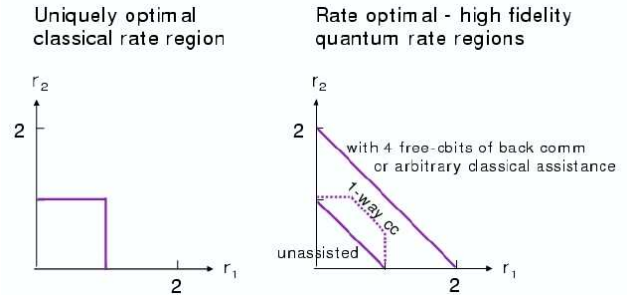}
\caption{Summary of the achievable rate regions of the butterfly
         network.  The entanglement assisted quantum rate region is
         also given by the left diagram.}
\label{fig:bf-sum}
\end{figure}

Since the inner and outer bounds are matching, the inner bound gives
the exact rate region (see \fig{bf-sum}), and the protocol described
is optimal.

$\bullet$ {\bf Two-way assisted case (with free two-way classical
communication)}

Note that the outer bound for the back-assisted case still applies,
thus, free two-way classical communication is no better than free back
classical communication alone.


$\bullet$ {\bf Forward-assisted case (with free forward classical
communication)}

Intriguingly, we will see how free forward classical communication can
effectively reverse the direction of some of the channels, but not all
of them.  Thus, the situation is intermediate between the unassisted
and the back-assisted cases.  We first describe a concrete
protocol for the butterfly network, before abstracting a general rule.

\begin{figure}[h]
\includegraphics[width=2in]{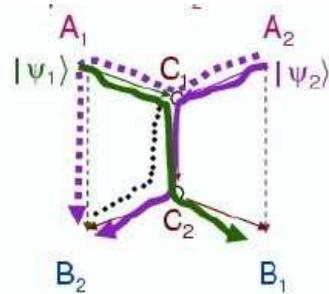}
 \caption{The achieving protocol for the rate pair $(1/2,1)$ in the
 back-assisted butterfly network.  The dotted paths represent the
 teleportation of $1$ qubit from $A_2$ to $B_2$, with the quantum
 portion sent via the thick dotted (purple) path, and the classical
 portion sent via the thin dotted (black) path.  Two other qubits are
 sent in the usual way via the other two solid paths.}
\label{fig:bf-q-fway}
\end{figure}

{\bf Inner bound:} The rate pair $(r_1,r_2) = (1/2,1)$ is achieved by
an exact, $2$-shot, protocol.  In the first network call, $A_1$
distributes $1$ ebit between $C_1$ and $B_2$.  $A_2$ sends one qubit
to $C_1$ who then teleports it to $B_2$.  Note that the classical
communication for the teleportation is sent via the path $C_1
\ra C_2 \ra B_2$.  (See the dotted path in Fig.\ \ref{fig:bf-q-fway}.) 
This leaves the $C_1 \ra C_2$ channel unused, leaving it as an
additional resource for the second network call.  For the second
network call, the two paths $A_1 \ra C_1 \ra C_2 \ra B_2$ and $A_2 \ra
C_1 \ra C_2 \ra B_1$ are used to communicate one qubit each from $A_1$
to $B_1$ and from $A_2$ to $B_2$.  These two paths are edge-avoiding
except for the $C_1 \ra C_2$ channel, but an additional use can be
borrowed from the first network call.  (See the solid paths in Fig.\
\ref{fig:bf-q-fway}).
Likewise, $(r_1,r_2) = (1,1/2)$ is also achievable. 
By monotonicity, $(1,0),(0,1)$ are also achievable, and so is the
convex hull of $(0,0),(1,0),(0,1),(1,1/2),(1/2,1)$. (See
\fig{bf-sum}.)


{\bf Outer bound:} To match the inner bound, we need to prove three
inequalities: $r_1,r_2 \leq 1$ and $r_1+r_2 \leq 3/2$.  Again, we
consider any $n$-use protocol communicating the $n_i =
n(r_i+\delta_n)$-qubit state $|\psi_i\>$ from $A_i$ to $B_i$.

The free forward classical communication provides many other
possibilities for encoding -- now into both quantum and classical
shares.  The classical shares can be cloned and ``broadcast
downstream'' for free.

Consider the encoding by $A_1$.  The state $|\psi_1\>$ is encoded into
$4$ shares: a quantum share for each of $A_1$, $C_1$, and $B_2$, and a
common classical share for each of them.  The shares to $C_1$ form an
authorized set, and by item \ref{qssqc}(b), the quantum portion has at
least $n_1$ qubits.  But there are only $n$ qubit-channels from $A_1$
to $C_1$.  Thus, $n_1 \leq n$ and $r_1 \leq 1$.  Similarly, $r_2 \leq
1$.

We now prove that $r_1+r_2 \leq 3/2$.  In particular, for any valid
protocol, consider using it in the following way: $A_1$ trying to send
$M_1$ where $M_1 R_1$ are prepared in $n (r_1 - \delta_n)$ ebits.
Similarly for $A_2$ and $R_2 M_2$.
But as mentioned before, $C_1$ received shares from $A_{1,2}$ that are
authorized for $M_{1,2}$ respectively.
This allows $C_1$ to play the ``man in the middle'' attack -- {\em to
keep} $M_{1,2}$ and {\em to replace} them by $M_{1,2}'$ maximally
entangled with $R_{1,2}'$ in his possession (i.e., he pretends to be
$B_{1,2}$ on the receiving end, and $A_{1,2}$ on the retransmitting
end).  (More formally, treat the entire $M_1 M_2 M_1' M_2'$ as the
combined quantum secret, and $C_1$ clearly holds an authorized set, 
and by item \ref{qssqc}(c) he can perform the logical swap between 
$M_1 M_2$ and $M_1' M_2'$.)  
A protocol with rate pair $(r_1,r_2)$ can then be modified to one that
shares $n (r_1 - \delta_n)$ ebits between $C_1$ and each of $A_1$ and
$B_1$, and $n (r_2 - \delta_n)$ ebits between $C_1$ and each of $A_2$
and $B_2$.  But only $3n$ qubits have gone in and out of $C_1$'s
laboratory in this modified protocol, upper bounding his total
entanglement with $A_{1,2} B_{1,2}$, which is $2n(r_1+r_2
-2\delta_n)$.  Thus, $r_1+r_2 \leq 3/2$, matching our inner bound for
the achievable rate region (see \fig{bf-sum}).

{\bf A general rule for reversing channels in forward-assisted 
quantum networks}  

\begin{figure}[h]
\includegraphics[width=2.5in]{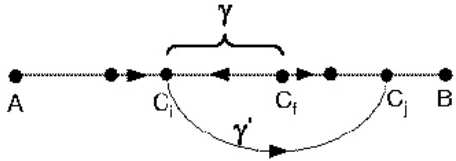}
 \caption{A sufficient condition for reversing an opposite running
          segment $\gamma$ in a communication path $\Gamma$ from $A$
          to $B$}
\label{fig:reversal-rule}
\end{figure}

Consider a path $\Gamma : A \leftrightarrow C_1 \leftrightarrow \cdots
\leftrightarrow B$, where $A,B$ are the sender and the receiver of
interest, $C_i$'s are intermediate parties, and the directions of the
quantum channels are variables in the problem.
For the purpose of $A \ra B$ quantum communication via $\Gamma$,
naively, we want the entire path to consist of forward channels, but
this turns out unnecessary.  
We state the following sufficient condition: 
\begin{quote}
The path $\Gamma$ from $A$ to $B$ can be used to communicate $1$ qubit
in a forward-assisted network if the following condition holds.
For each segment $\gamma$ of $\Gamma$ running in the opposite
direction, with boundary points $C_{i},C_{f}$, there is an entirely
forward path $\gamma'$ in the network from $C_{i}$ to some $C_j \in
\Gamma$ with $j \geq f$.  Besides the boundary points, $\gamma$ and
$\Gamma$ impose no further constraint on $\gamma'$.
\end{quote}
In other words, an opposite running segment poses no problem as long
as the network provides some forward path bridging its beginning to
its end or beyond (see \fig{reversal-rule}). 
To prove the sufficiency of this condition, we use teleportation.  The
opposite running segment $\gamma$, together with the segment between
$C_f$ and $C_j$, can be used to establish an ebit between $C_i$ and
$C_j$. $C_i$ then teleports the message to $C_j$.

$\bullet$ {\bf Entanglement-assisted case} 

We first define the assisting resource.  Here, we assume that any two
parties share free ebits.  We discuss alternative models later.

Since $A_1, B_1$ share ebits, and similarly for $A_2, B_2$, by
teleportation and superdense-coding, the rates for quantum
communication are exactly half of those for classical communication
via the quantum network, so we focus on the latter.

{\bf Inner bound for classical communication:} 
Given free ebits, each quantum channel in the network can transmit $2$
cbits by superdense coding \cite{BW92}.  Twice the unit square is
achievable.

{\bf Outer bound for classical communication:} 
The Holevo bound \cite{Holevo73} (see also \cite{CDNT97}) states that
by using $n$ forward qubit-channels, unlimited back quantum
communication, and arbitrary prior entanglement 
one cannot send more than $2n$ forward cbits. 
Consider the cut $S_1 = \{A_1,B_2\}$ and $S_2 = \{A_2, B_1, C_1,
C_2\}$.  Since there is only one forward quantum channel, no more than
$2$ cbits can be communicated from $A_1$ to $B_1$ per use of the
network.  Similarly for the classical communication from $A_2$ to
$B_2$.  

Thus the exact rate region for classical communication is twice 
the unit square, and that for quantum communication is the unit 
square.  

{\bf Alternative assisting models}

Another natural model of assistance is to allow free ebits only
between neighboring parties in the network.
We leave the achievable rate region for the butterfly network in this
case as an open question.  So far, we cannot find a good protocol that
achieves the quantum rate pair $(1,1)$.  We believe that it is not
achievable.  If our belief holds, this alternative model has a
continuity problem.
Consider adding a complete graph of channels to the network, with
arbitrarily small capacity for each edge, so that all pairs of parties
are now ``neighbors.''  Then, the pair $(1,1)$ is achievable -- these
added channels with negligible capacities change the communication
rates abruptly.

Since the initial posting of this manuscript, Hayashi \cite{H07} has
considered entanglement assistance between the senders and also
between neighbors in the network, but this model is out of our present
scope.

\vspace*{1ex} 

$\bullet$ {\bf Summary for the butterfly network}

%


\section{Another example - the inverted crown network} 
\label{sec:examples}

\begin{figure}[h]
\includegraphics[width=1.6in]{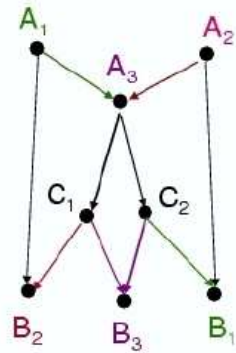}
 \caption{The inverted crown network}
\label{fig:icrown}
\end{figure}

We consider the quantum version of a more complicated network studied
in \cite{HKL04} to illustrate our more general results.  It is depicted
in \fig{icrown} and we will call it the inverted crown network.

We will use techniques similar to those in \msec{butterfly}, skipping
details in the arguments that should now be familiar.

$\bullet$ {\bf Unassisted case}

{\bf Inner bound:} 
The rate triplets $(1,1,0)$, $(0,0,2)$, $(1,0,1)$, and
$(0,1,1)$ are achievable due to the following sets of paths:
\begin{figure}[h]
\includegraphics[width=2.0in]{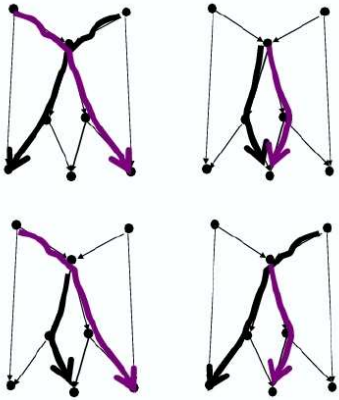}
 \caption{Paths for achieving the extremal rate triplets $(1,1,0)$,
 $(0,0,2)$, $(1,0,1)$, and $(0,1,1)$.}
\label{fig:icrown0}
\end{figure}

By monotonicity, $(1,0,0)$ and $(0,1,0)$ are also achievable.  The 
convex hull of these points (together with the origin) is plotted in 
\fig{icunassist}.

{\bf Outer bound:} 
We will first use the mincut method (item \ref{min} in
\msec{butterfly}).  Let $S_{1,2}$ be a bipartite cut.  Consider
forward communication from $S_1$ to $S_2$.  We obtain the following
bounds:
\bea
 r_1 \leq 1 & {\rm for} & S_1 = \{A_1,C_1,B_2,B_3\}
\nonumber
\\
 r_2 \leq 1 & {\rm for} & S_1 = \{A_2,C_2,B_1,B_3\}
\nonumber
\\
 r_3 \leq 2 & {\rm for} & S_2 = \{B_3\}
\nonumber
\\
r_1 + r_2 \leq 2 & {\rm for} & S_1 = \{A_1,B_2\}
\nonumber
\\
r_1 + r_3 \leq 2 & {\rm for} & S_1 = \{A_1,A_3,C_1,B_2\}
\label{eq:icumanybdds}
\eea
Note that the $3$rd and the $4$th bounds hold even with free two-way
classical communication.

By inspection, the inner bound in \fig{icunassist} can be matched
given the $1$st and $2$nd inequalities above, together with
$r_1+r_2+r_3\leq 2$.
The last inequality can be proved similarly to the case for the
butterfly network, and we will be brief here.  Let $M_i$ be the
$n_i$-qubit message from $A_i$ to $B_i$, with reference $R_i$.
We take the quantum secret to be $M_1,M_2,M_3$, and the combined
reference $R=R_1,R_2,R_3$.  
Let 
$Q_{1}$ denote the $A_1 \ra B_2$ communication, 
$Q_{2}$ the $A_2 \ra B_1$ communication, and 
$Q_{C}$ denote the $A_3 \ra C_{1,2}$ communications combined. 
Again, for $B_1$ to recover $M_1$, $I(Q_1\,{:}\,R_1)$ has to be 
small, and similarly for $I(Q_2\,{:}\,R_2)$. 
By causality $R_1 Q_1$, $R_2 Q_2$ and $R_3$ are all independent.
Then, $I(Q_{1} Q_2 \,{:} \, R) {:}{=} S(Q_1 Q_2) + S(R_1 R_2 R_3) -
S(Q_1 Q_2 R_1 R_2 R_3) = I(Q_1\,{:}\,R_1) + I(Q_2\,{:}\,R_2)$ is
small, where once again, the equality is due to the various
independence conditions.
But $Q_{1,2,C}$ is authorized, thus $Q_C$ is significant, and has at
least $n_1+n_2+n_3$ qubits, while having at most $2n$ qubits.  Thus
$r_1+r_2+r_3 \leq 2$ as claimed, and the inner bound is matched by the
outer bound.

We remark that in the analoguous problem of sending classical
information through the classical inverted crown network, the same
outer bound on the rate region holds.  (Bounds on $r_{1,2}$ due to the
mincut property also hold classically, and \cite{HKL04} proves that 
$r_1+r_2+r_3 \leq 2$.)

$\bullet$ {\bf Forward-assisted case}

\begin{figure}[h]
\includegraphics[width=2.4in]{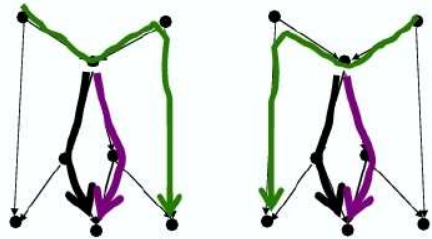}
 \caption{Sets of paths for achieving the extremal rate triplets
 $(1,0,2)$ and $(0,1,2)$ for the forward-assisted inverted crown
 network.}
\label{fig:icrown1}
\end{figure}

{\bf Inner bound:} 
The rate triplets $(1,1,0)$, $(1,0,2)$, $(0,1,2)$ are achievable.  The
first is achieved without assistance (see previous subsection).  The
point $(1,0,2)$ is achieved by the paths depicted in the left diagram
of \fig{icrown1}.  To reverse the path $A_2 \ra A_3$, we use the
``bridge'' $\gamma' = A_3 \ra C_2 \ra B_1$ (see the general rule for
reversing paths in \msec{qbutterfly}).  Similarly for the triplet 
$(0,1,2)$.
Thus we obtain an inner bound that is the convex hull of $(0,0,2)$,
$(1,1,0)$, $(1,0,2)$, $(0,1,2)$, $(1,0,0)$, $(0,1,0)$, and the origin.
(See \fig{icunassist}.)

{\bf Outer bound:} 
{From} \fig{icunassist}, it suffices to show that $r_{1,2} \leq 1$,
$r_3 \leq 2$, and $2(r_1+r_2) + r_3 \leq 4$ in order to match the
inner bound.
We have $r_{1,2} \leq 1$ even with free forward classical
communication, because the messages to $A_3$ still form an authorized
set and quantum-classical compression does not decrease the sizes of
the quantum parts.  The bound $r_3 \leq 2$ proved in the unassisted
case holds even with two-way assistance.
The remaining bound $2(r_1+r_2) + r_3 \leq 4$ can be proved as
follows.

In the absence of back communication, the $A_1 \ra A_3$ message has to
be authorized for $|\psi_1\>$ and the $A_2 \ra A_3$ message has to be
authorized for $|\psi_2\>$.  Running an argument similar to that for
the butterfly network, on $A_3$ replacing $|\psi_{1,2}\>$ by her own
messages, a protocol that communicates $n_i$ qubits from $A_i$ to
$B_i$ can be used to establish $2 (n_1 +n_2) + n_3$ ebits between 
$A_3$ and $A_{1,2} B_{1,2,3}$.  But there are only $4n$ channels in and
out of $A_3$, thus $2(r_1+r_2) + r_3 \leq 4$.

We summarize the results for the last two subsections in the following 
figure: 
\begin{figure}[h]
\includegraphics[width=3.4in]{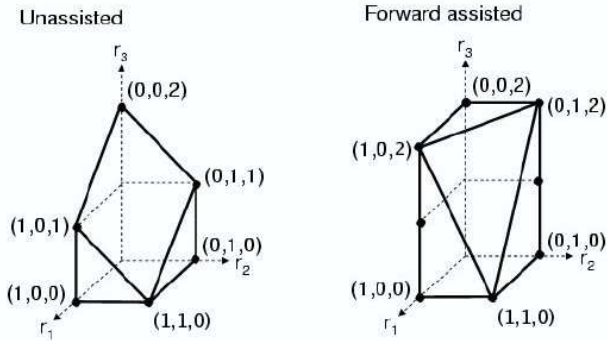}
 \caption{The achievable rate regions for the inverted crown network
 in the cases with no free classical communication and free forward
 classical communication.}
\label{fig:icunassist}
\end{figure}

$\bullet$ {\bf Back-assisted case}

{\bf Inner bound} 

\begin{figure}[h]
\includegraphics[width=2.4in]{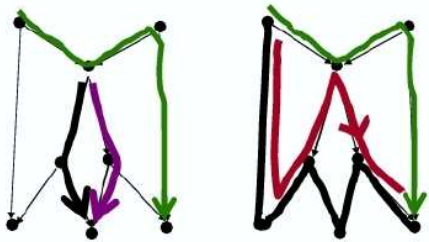}
 \caption{Paths for achieving the extremal rate triplet $(2,0,1)$ for
 the backward or two-way assisted inverted crown network.  All six
 paths can be used by calling the network twice, thus achieving the
 stated rates.}
\label{fig:icrown2}
\end{figure}

\begin{figure}[h]
\includegraphics[width=2.4in]{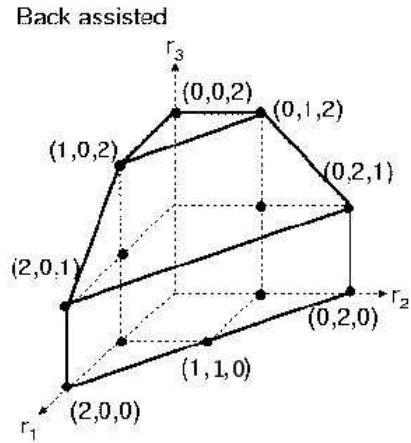}
 \caption{The achievable rate region for the inverted crown network
 given free backward or two-way classical communication.}
\label{fig:icbackassist}
\end{figure}

The rate points $(1,0,2)$ and $(0,1,2)$ can be achieved as in the
forward-assisted case.  In addition, the rate points $(2,0,1)$ and
$(0,2,1)$ are also achievable.  The paths to achieve the former are
shown in \fig{icrown2}, and the latter can be achieved similarly.
Thus, we obtain \fig{icbackassist} for the inner bound.  

{\bf Outer bound} 

Consider the cuts used in \eq{icumanybdds} for the unassisted case,
but allow free two-way classical communication now.  The 3rd and the
4th inequalities, $r_3 \leq 2$ and $r_1+r_2 \leq 2$, stay the same,
while the 5th inequality becomes $r_1+r_2+r_3 \leq 3$, matching our
inner bound.
%
%
%

$\bullet$ {\bf Entanglement assisted case}
 
Inner bounds can be obtained from known classical solutions
\cite{HKL04}, whose outer bounds will also apply if proposition {\bf
P} in \msec{eassist} is proven true.

\section{Generalization to other networks} 
\label{sec:general}

As we have seen in \msec{butterfly}, routing (with time sharing) is
sufficient to generate the entire achievable rate region.  It is
suggestive that routing is indeed optimal for more general networks,
and finding maximal sets of edge avoiding paths provides optimal
protocols.  We have not been able to prove such a conjecture in full
generality.  In this section, we present some ideas and proofs in 
special cases.

In the following, each channel in the network has a capacity that is
an arbitrary nonnegative number. Since we allow an asymptotically
large number of network calls, without loss of generality, the
capacities can be taken as integers.  Conditions imposed on the
network and assisting resources will vary from case to case.

We have not encountered a situation that requires nontrivial
time-ordering of individual channel uses (within or across network
calls) to achieve optimality.  
In any case, time-ordering will not affect the optimal rates since in
the asymptotic limit, nontrivial time-ordering can be effectively
achieved, by using a negligible fraction of earlier network calls
inefficiently or by ``double-blocking'' (running in parallel many
copies of an arbitrarily ordered $n$-use protocol).

\subsection{The case with a general number of sender-receiver 
pairs in shallow networks}
\label{sec:shallow}

\begin{figure}[h]
\includegraphics[width=3in]{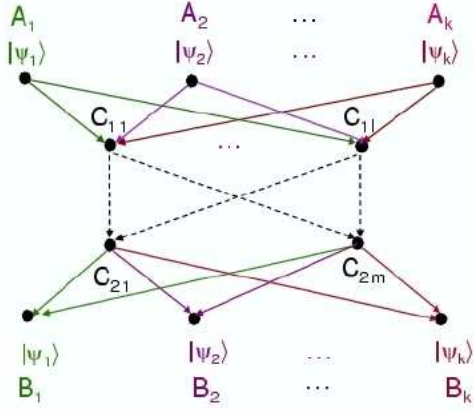}
 \caption{A general shallow network of distance $d$.}
\label{fig:shallow}
\end{figure}

\begin{figure}[h]
\includegraphics[width=1.8in]{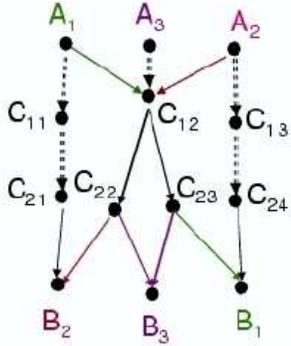}
 \caption{The inverted crown network in the form of \fig{shallow}. 
          The dotted channels have unlimited capacities.}
\label{fig:icrownrewrite}
\end{figure}

In this class of networks, we impose three conditions: (1) there are
no out-going channels from any receiver in the given network, and (2)
the maximum length of any simple (i.e.\ without closed loops) path
from a sender to any receiver is bounded by $d=3$, and (3) there is no
$4$-cycle involving a sender (detail later).
The most general situation manifesting conditions (1) and (2) is
depicted in \fig{shallow}, 
\newline and condition (3) disallows any $4$-cycle with vertices $A_i,
C_{1j_1}, C_{2j_2}, C_{1j_3}$ for any $ij_1j_2j_3$.
For an arbitrary positive integer $k$, for each $i = 1,\cdots,k$, a
sender $A_i$ wants to send a message $|\psi_i\>$ to the receiver
$B_i$.
It is crucial that $|\psi_i\>$ are independent messages. 
There are outgoing channels from the $A_i$'s to the $C_{1j}$'s
($j=1,\cdots,l$), from the $C_{1j}$'s to the $C_{2j}$'s
($j=1,\cdots,m$), and from the $C_{2j}$'s to the $B_i$'s.  The absence
of a channel is signified by a zero capacity.
Note that elements in the sets $\{A_i\}$, $\{C_{1,j}\}$, and
$\{C_{2,j}\}$ are defined by their distances to the $A_i$'s
and $B_i$'s.
In a general network, these three sets need not be disjoint.  For
example, a sender $A_i$ may receive information from others and may
directly communicate with a specific $B_j$.  
The first case holds for $A_3$ in the inverted crown network in 
\fig{icrown} and the second case holds for both $A_{1,2}$ in the 
butterfly network in \fig{bf-setup}.  
Such a configuration can easily be handled by assigning multiple
vertices to the same party (e.g.\ $A_i$ is duplicated as an additional
$C_{1,j}$) and connecting the parties with a high capacity channel
(from $A_i$ to $C_{1,j}$ in this example).
Thus, Fig.\ \ref{fig:shallow} still covers these cases.
To illustrate the idea, we express the inverted crown network
(\fig{icrown}) in the form of \fig{shallow} in \fig{icrownrewrite}.

{\bf Unassisted case:} 

We show that routing is optimal in the case without any classical
communication assistance.

{\bf Proof} (or proof ideas) {\bf:}

Consider the most general $n$-use protocol.  For each channel, we can
group together the messages from all $n$ uses as a single piece.

Denote the message from $C_{2j}$ to $B_i$ as $Q_{2ji}$.  
The quantum messages $\{Q_{2ji}\}_{j}$ received by $B_i$
form an authorized set for $|\psi_i\>$.
It also means that the quantum messages $\{Q_{2ji'}\}_{j i'\neq i}$
form an unauthorized set.
In other words, the entire set of messages $\{Q_{2ji}\}_{ji}$ from all
the $C_{2j}$'s to all the $B_i$'s form a tensor product encoding scheme
for the tensor product secret $|\psi_1\> \ot \cdots \ot |\psi_k\>$.
The salient property here is that, we can say that message $Q_{2ji}$ 
is a share of $|\psi_i\>$ alone, and not of any other $|\psi_{i'}\>$ 
for $i'\neq i$. 
(Possible entanglement between the messages of $B_i$ and $B_j$ will be
decoded to a state independent of their messages $|\psi_i\>$ and
$|\psi_j\>$.)

Now consider messages from the $C_{1j}$'s to the $C_{2j}$'s.  Each
$C_{1j}$ has just received from each $A_i$ and if he entangles the
messages from $A_{i_1}$ and $A_{i_2}$ and distributes shares to
different $C_{2\tilde{j}}$'s, the latter will not be able to re-encode
the shares to the product form mentioned above (the only case in which
this is not obvious is a $4$-cycle of the forbidden type).
Thus, the messages $|\psi_i\>$ are never jointly coded in any part of
the network, and the optimality of routing follows.

Our proof technique has not taken advantage of the optimality of the
protocol analyzed.  In the presence of these $4$-cycles, we cannot
rule out a protocol that entangles the messages at the $C_{1j}$'s, but
such a protocol appears less efficient.  Unfortunately, we have not
been able to turn this intuition into a rigorous argument.

{\bf Forward-assisted, back-assisted and two-way assisted cases:} 

The forward-assisted case can be analyzed similarly, provided the
possible inversion of the intermediate edges is taken into account.
In the back-assisted case, and in network not obeying condition (1),
the receivers are no longer information sinks, but it still holds that
each receiver can only {\em retain} unauthorized shares of other
messages (though they can help in transmitting them).
However, including all the extra possible paths through the receivers 
makes most networks too deep for the proof to apply.
For example, our proof applies to the butterfly network
(\fig{bf-setup}) but not the inverted crown network (\fig{icrown}).

{\bf General discussion} 

The proof ideas used in this subsection, unfortunately, do not extend
readily to deeper networks.  Whether there are deeper networks that
require entangling coding strategies remains an interesting open issue
to be resolved.

\subsection{Outer and inner bounds on the achievable rate regions} 
\label{sec:mincut} 

Consider the $k$-pair communication problem in the most general
network.  For any subset $\Sigma$ of the $k$ pairs of sender/receiver,
we will derive upper and lower bounds for their rate sum.

{\bf Outer bound} 

The upper bound of the rate sum is via the min-cut idea discussed at
the beginning of \msec{butterfly}.
Let $S$, $R$ be any bipartite cut (partition) of the vertices such
that the senders of $\Sigma$ are in $S$ and the receivers are in $R$.
Let $c_\ra(S)$ be the sum of the capacities of all the channels from
$S$ to $R$ and $c_\la(S)$ be that from $R$ to $S$.
Then, the rate sum for $\Sigma$ is upper bounded by $\min_S c_\ra(S)$
in the unassisted case.
A weaker bound holds in networks assisted by forward, backward, or
two-way classical communication as follows.
For any cut $S$, $R$ that separates each sender/receiver pair in
$\Sigma$, the rate sum is upper bounded by $\min_S c_\ra(S)+c_\la(S)$.
%
%
%
%
To see the first statement, for any cut $S$ and $R$, any protocol for
the $k$-pair communication problem gives a method to communicate from
$S$ to $R$, whose rate cannot exceed $c_\ra(S)$.  For the second
statement, any protocol for the $k$-pair communication problem gives a
method to generate entanglement between $S$ and $R$ at a rate that
cannot exceed $c_\ra(S)+c_\la(S)$ even when assisted by free two-way
classical communication.

{\bf Inner bound} 

A lower bound for the rate sum for $\Sigma$ is given by constructing
edge avoiding paths.  Here, we can interpret an edge with capacity $c$
as $c$ edges of unit capacity.  (Recall that integer values of
capacities are general.)  The lower bound for the rate sum is simply
the maximum number of paths connecting each sender in $\Sigma$ to the
correct receiver.  In the unassisted case, all edges in each path have
to be properly oriented; similarly in the forward assisted case,
except we allow reversal of the edges if the general rule described in
\msec{butterfly} is satisfied; in the back assisted or two-way
assisted case, the edges are simply undirected.

\subsection{$2$-pair communication in arbitrary networks with 
back-assistance} 
\label{sec:ghost}

The setting is a special case of the previous subsection with $k=2$
and with free classical back communication (thus the channels are
undirected).  Here, we will first tighten the rate sum.  Then, we show
that the upper bounds on the individual rates and the rate sum
completely define the achievable rate region, by proving their
achievability.

{\bf Improved upper bound on the rate sum} 

Consider any $n$-use protocol that communicates from $A_1$ to $B_1$
and from $A_2$ to $B_2$ with a rate sum $r$.  
Clearly the protocol can generate, at the same rate, entanglement
between $A_1,A_2$ and $B_1,B_2$.
Thus, for any bipartite cut $S_h,R_h$ separating $A_1,A_2$ from
$B_1,B_2$, if $c_\leftrightarrow(S_v)$ is the sum of the capacities of
all the channels (in both directions) between $S_v$ and $R_v$, then,
$r \leq c_\leftrightarrow(S_v)$.
But the communication protocol also generates entanglement between
$A_1,B_2$ and $A_2,B_1$ at a rate $r$, and applying an argument
similar to the above, $r \leq c_\leftrightarrow(S_h)$ for any
bipartite cut $S_h,R_h$ separating $A_1,B_2$ from $A_2,B_1$.
Minimizing over all $S_v$, $S_h$, it follows that 
\be 
	r \leq \min \left[ \min_{S_v} c_\leftrightarrow(S_v) \, , \, 
	  \min_{S_h} c_\leftrightarrow(S_h) \right] \,.
\label{eq:2pair-ratesum}
\ee

{\bf Achievability} 

\begin{figure}[h]
\includegraphics[width=2.5in]{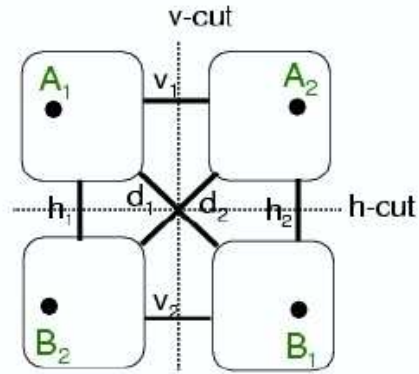}
 \caption{How the two mincuts partition the vertices into $4$ groups}
\label{fig:2pairs}
\end{figure}

We now show that the above upper bound on the rate sum, together with
the upper bounds on the individual rates given by \msec{mincut}, define
the achievable rate region.
Take the partitions $S_v,R_v$ and $S_h,R_h$ that respectively minimize
$c_\leftrightarrow(S_v)$ and $c_\leftrightarrow(S_h)$, and take
intersections to obtain a partition of the vertices into $4$ subsets
(see \fig{2pairs}).
We can bundle the channels between these $4$ subsets into $6$ groups
(labeled $v_{1,2}, h_{1,2}$, and $d_{1,2}$ pertaining to the vertical
cut, the horizontal cut, and the diagonals).

\begin{figure}[h]
\includegraphics[width=2.5in]{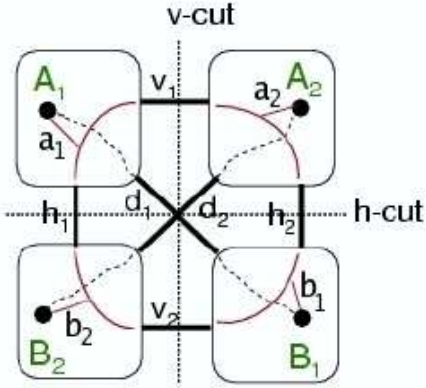}
 \caption{The structure of an arbitrary quantum network relevant to
 the $2$-pair communication problem with free back communication}.
\label{fig:2pairsproof}
\end{figure}

Use the max-flow-min-cut theorem (see \msec{butterfly}), we can find
the paths for the vertical cut, extending $v_{1,2}, d_{1,2}$ to
$A_{1,2}, B_{1,2}$, and similarly for the horizontal cut.
(See the red paths in \fig{2pairsproof}).
%
%
%
%
%
%
%
%
The paths through $d_{1,2}$ avoid all other red paths, but those
through $h_{1,2}$ and $v_{1,2}$ may share edges.  
In \fig{2pairsproof} we schematically show merged paths running
towards $A_{1,2}, B_{1,2}$.  Most generally, the red paths may merge
and diverge in other locations, but the important features are that
they reach $A_{1,2}$ and $B_{1,2}$, and may impose bottlenecks for
flows in/out of the individual $A_{1,2},B_{1,2}$.  We label the
possible bottlenecks by $a_{1,2},b_{1,2}$.  With a slight abuse of
notations, we denote the capacities of $a_1,a_2, \cdots, h_2$ by the
same symbols.  Then,
\bea
	r_1 & \leq & r_1^* \; {:}{=} \; \min(a_1,b_1,v_1{+}h_1,v_2{+}h_2)+d_1
\\
	r_2 & \leq & r_2^* \; {:}{=} \; \min(a_2,b_2,v_1{+}h_2,v_2{+}h_1)+d_2
\\
	r & \leq & r^* \; {:}{=} \; \min(v_1{+}v_2,h_1{+}h_2)+d_1+d_2
\label{eq:ratesum}
\eea

%
To achieve the rate pair $(r_1^*,r^*-r_1^*)$, $A_1$ and $A_2$ use the
paths $d_{1,2}$ independently.
They have to share the use of $v_1,h_1,v_2,h_2$ (collectively called 
the ``square'').
$A_1$ can send qubits independently through the paths 
\bea
	\gamma_1 & : & a_1 \ra v_1 \ra h_2 \ra b_1
\\	\gamma_2 & : & a_1 \ra h_1 \ra v_2 \ra b_1
\eea
{\bf Case (1)} If $r_1$ is limited by $a_1$ or $b_1$, $A_1$ sends
${1\over 2} \min(a_1,b_1)$ qubits through each of $\gamma_{1,2}$.
But the rate sum is limited by the square, and it is easy to check
that the unused channels in the square support enough $A_2 \ra B_2$
communication to achieve the rate sum given by \eq{ratesum}.
{\bf Case (2)} If $r_1$ is limited by the square, $A_1$ sends
$\min(v_1,h_2)$ qubits through $\gamma_1$ and $\min(h_1,v_2)$ qubits
through $\gamma_2$.
{\bf Case (2a)} If $v_1 < h_2$ and $h_1 < v_2$, the path $h_2 \ra v_2$
will be available for $A_2$ to communicate to $B_2$ to achieve the
rate sum.
Similarly for the case $v_1 > h_2$ and $h_1 > v_2$. 
{\bf Case (2b)} Otherwise, either $h_1+h_2$ or $v_1+v_2$ will be the
limiting factor, and the rate sum is already achieved by maximizing
$r_1$ in the way described above, without the need for any further
contribution from the $A_2 \ra B_2$ communication.

By symmetry of the problem, the rate pair $(r^*-r_2^*,r_2^*)$ is also
achievable.  Invoking monotonicity and time sharing, the
characterization of the achievable region is completed.




\subsection{Any arbitrary entanglement-assisted network}
\label{sec:eassist} 

Our discussion for the quantum butterfly network applies to the most
general quantum network communication problem.  Because of superdense
coding and teleportation, the achievable rate region for classical
communication in a quantum network is exactly twice of that for
quantum communication.  The latter is clearly inner bounded by the
achievable rate region for sending classical data via the
corresponding classical network.
This inner bound is tight if the following network generalization of
Holevo's bound holds.

Let {\bf P} be the following proposition:
\begin{quote}
If $(r_1,r_2,\cdots)$ is not an achievable point in the classical rate
region of a classical network, then, $(2r_1,2r_2,\cdots)$ is not an
achievable point in the classical rate region of the corresponding
quantum network with arbitrary entanglement assistance.
\end{quote} 
If proposition {\bf P} holds, then, the exact achievable rate
region for quantum (classical) communication in an
entanglement-assisted quantum network is exactly (twice) the classical
rate region of the (unassisted) classical network.

\section{Other network communication problems} 
\label{sec:others}

In this section, we discuss two other quantum network communication
problems that are very different from the $k$-pair communication
problem.

\subsection{Quantum multicasting}
\label{sec:multicast}

Reference \cite{ACLY00} also studies the {\em multicast problem} in
which a single source transmits the same message to $k$ different
receivers.

We define a quantum analogue to the problem, by considering $k$ pairs
of senders and receivers, $A_i$ and $B_i$.  A reference party $R$
creates the state $|\psi\> = {1 \over \sqrt{2}} (|0\>^{\ot (k{+}1)} +
|1\>^{\ot (k{+}1)})$ and gives one qubit of $|\psi\>$ to each $A_i$,
keeping one qubit to himself.  The goal is for $R$ to share $|\psi\>$
with the $B_i$'s, enabled by the quantum communication through the
given quantum network.  The optimal rate is given by the maximum
number of copies of $|\psi\>$ shared per use of the network, allowing
a large number of network calls.

In the quantum problem, one can achieve at least the rate region of
the classical problem, by applying any classical strategy in the
computation basis.  Whether this inner bound is tight or not is an
open problem.

\subsection{Multi-party entanglement of assistance}
\label{sec:EoA}

In quantum communication theory, in some settings, one believes that
it is easy to perform classical communication but hard to obtain any
quantum resource.
It is a common scenario to assume that the remote parties share a
quantum state and the problem is to determine the amount of quantum
communication that can be generated given unlimited classical
communication (between all of them).
By teleportation, the problem reduces to generating ebits between
sender-receiver pairs.  Much has been done for one sender-receiver
pair, analoguous to the one sender-receiver pair situation in network
communication.
Here, we consider the problem of generating ebits simultaneously among
many pairs of parties, which relates to simultaneous network
communication (though not specifically the $k$-pair communication
problem).
Related problems have been considered recently
\cite{SVW05,HOW05s,HOW05l,FL06}.

Suppose $m$ parties share a pure state $|\psi\>$.  Parties $A_1,B_1$
are special.
The other $m{-}2$ parties are allowed to send classical communication to
them (but not vice versa).
The entanglement of assistance for $A_1,B_1$~\cite{DFMSTU98}, (also
known as localizable entanglement~\cite{PVMC04}) is defined as the
maximum number of ebits they can share afterwards.
Clearly, the optimal strategies for those $m{-}2$ parties are to make
measurements (with rank-$1$ measurement operators) and to communicate
the measurement outcomes to $A_1,B_1$.
%
%
This gives rise to the following expression for the {\em regularized}
entanglement of assistance, the maximum number of ebits between
$A_1,B_1$ created per copy of the state $|\psi\>$, when large number
of copies are shared.
\be \eass^\infty (|\psi\>,A_1{:}B_1) = \sup \sum_k p_k E(|\psi_k\>) \ee
where the supremum is taken over the $m{-}2$ local measurements, $k$
denotes the $m{-}2$ measurement outcomes, and $|\psi_k\>$ is the
corresponding postmeasurement state of $A_1,B_1$, and $E(\cdot)$ is
the usual measure for pure state bipartite entanglement defined as
follows.  For any pure bipartite state $|\phi\>$, let $\phi_{1,2}$ be
the reduced density matrices on the two parties and $S(\rho):= -\tr
\rho \log \rho$ be the von Neumann entropy of $\rho$.  Then,
$E(|\phi\>) := S(\phi_{1}) = S(\phi_{2})$.  Thus, for a set of parties
holding a pure state and a subset of parties $\Sigma$, we simply write
$S(\Sigma)$ for the entanglement between $\Sigma$ and the rest of the
parties.
It was found in \cite{SVW05,HOW05l} that 
\beq
  \eass^\infty (|\psi\>,A_1{:}B_1) 
  = \min_{T} \{ S(AT) , S(BT^c) \}
\eeq
where $T$ and $T^c$ is a partition of the other $m{-}2$ parties.  
With extra classical communication from $A_1$ to $B_1$, we relate back
to the usual communication problem of one sender-receiver pair in a
{\em static version} of a network (a multipartite state).

In fact, by the state merging protocol in \cite{HOW05s,HOW05l}, for
{\it any} $T,T^c$, each copy of $|\psi\>$ can generate: \\
(1) $S(A_1 T)$ ebits between $A_1$ and $B_1$ \\
(2) $-S(T|A_1)$ ebits between $T$ and $A_1$ \\
(3) $-S(T^c|B_1)$ ebits between $T^c$ and $B_1$ \\
where the conditional entropy $S(T|A_1$) is defined as $S(A_1
T)-S(A_1)$ and similarly for $-S(T^c|B_1)$.

In fact, using the chain rule, one can see that each party within the
groups $T$ and $T^c$ can distill (or consume) an amount of
entanglement given by $-S(T_i|A_1T_1T_2...T_{i-1})$ for the parties in
the partition $T$ and likewise $-S(T^c_i|B_1T^c_1T^c_2...T^c_{i-1})$
for the individual parties within $T^c$.  Here, the $T_i$ are the
parties in the partition $T$, and $T^c_i$ are in $T^c$.  Yang and
Eisert [arXiv:0907.4757] have since shown that if the partition $T^c$
is empty, then one can eliminate the need to consume entanglement.

This allows more sender-receiver pairs to communicate with one
another depending on the initial state and the exact form of 
classical communication assistance.  

\section{Conclusion}
\label{sec:conclusion}

We have studied the $k$-pair communication problem for quantum data in
quantum networks under different assisted scenarios.  We obtained a
general statement for the optimality of routing in shallow networks
without a certain type of $4$-cycles and worked out the exact rate
regions in a number of simple cases.  A number of problems remain
unresolved, including the validity of proposition {\bf P} in
\msec{eassist} (outer-bounding the entanglement assisted classical
rate points in quantum networks), and the optimality of routing in
networks with larger depth.

\begin{acknowledgments}

We are indebted to David Bacon for drawing our attention to reference
\cite{SS06} and Daniel Gottesman for discussions on quantum secret
sharing and for providing one of our upper bounds.  We thank Andris
Ambainis and Rahul Jain for comments on communication complexity,
Yaoyun Shi for constructive criticisms, and Michael Ben-Or and
Nicholas Harvey for chats on other open problems in the classical
setting.
We thank Panos Aliferis for an inspiring verse, which is
modified and shared here: 
\vspace*{-2ex} 
\begin{quote} 
Eat and drink and smoke and write
and sleep and eat and drink and play and prove and correct and laugh
and $\ldots$ submit!
\\
\hspace*{2ex} $\ldots$ and sleep and eat and edit and replace.
\end{quote} 

DL was supported by CRC, CRC-CFI, CIAR, NSERC, ARO, and MITACS.  JO
acknowledges the support of the Royal Society and EU Project QAP
(IST-3-015848).  AW was supported by EU grant RESQ (IST-2001-37559),
the U.K. Engineering and Physical Sciences Research Council's ``QIP
IRC'', and a University of Bristol Resesarch Fellowship.
\end{acknowledgments}


\begin{thebibliography}{10}

\bibitem{ACLY00}
R.~Ahlswede, N.~Cai, S.-Y.~R. Li, and R.~W. Yeung,
\newblock ``Network information flow,''
\newblock {\em IEEE Trans. Inf. Theory}, vol. 46, no. 4, pp. 1204--1216, 2000.

\bibitem{YZ99}
R.~Yeung and Z.~Zhang,
\newblock ``Distributed source coding for satellite communications,''
\newblock {\em IEEE Trans. Inf. Theory}, vol. 45, no. 4, pp. 1111, 1999.

\bibitem{LL04a}
Z.~Li and B.~Li,
\newblock ``Network coding: The case of multiple unicast sessions,''
\newblock in {\em In Proceedings of the 42nd Allerton Annual Conference on
  Communication, Control, and Computing}, 2004.

\bibitem{HKL04}
N.~J.~A. Harvey, R.~D. Kleinberg, and A.~R. Lehman,
\newblock ``Comparing network coding with multicommodity flow for the $k$-pairs
  communication problem,''
\newblock {\em Technical Report MIT-LCS-TR-964}, September 2004.

\bibitem{LL04b}
Z.~Li and B.~Li,
\newblock ``Network coding in undirected networks,''
\newblock in {\em In Proceedings of the 38th Conference on Information Science
  and Systems (CISS)}, 2004.

\bibitem{JVYY06}
K.~Jain, V.~Vazirani, R.~Yeung, and G.~Yuval,
\newblock ``On the capacity of multiple unicast sessions in unidirected
  graphs,''
\newblock {\em IEEE Trans. Inf. Theory}, vol. 52, no. 6, pp. 2805--2809, 2006.

\bibitem{HKL06}
N.~J.~A. Harvey, R.~D. Kleinberg, and A.~R. Lehman,
\newblock ``On the capacity of information networks,''
\newblock {\em IEEE Trans. Inf. Theory}, vol. XX, no. Y, pp. ZZZZ, 2006.

\bibitem{HINRY06}
M.~Hayashi, K.~Iwama, H.~Nishimura, R.~Raymond, and S.~Yamashita,
\newblock ``Quantum network coding,'' 2006,
\newblock quant-ph/0601088.

\bibitem{SS06}
Y.~Shi and E.~Soljanin,
\newblock ``On multicast in quantum networks,''
\newblock in {\em Proceedings of the 40th Annual Conference on Information
  Sciences and Systems (CISS)}, 2006.

\bibitem{LOW06}
D.~Leung, J.~Oppenheim, and A.~Winter,
\newblock ``Quantum network communication -- the butterfly and beyond,'' 2006,
\newblock quant-ph/0608223.

\bibitem{BBCJPW93}
C.~H. Bennett, G.~Brassard, C.~Cr\'epeau, R.~Jozsa, A.~Peres, and W.~Wootters,
\newblock ``Teleporting an unknown quantum state via dual classical and
  {E}instein-{P}odolsky-{R}osen channels,''
\newblock {\em Phys. Rev. Lett.}, vol. 70, pp. 1895--1899, 1993.

\bibitem{BW92}
C.~H. Bennett and S.~J. Wiesner,
\newblock ``Communication via one- and two-particle operators on
  {E}instein-{P}odolsky-{R}osen states,''
\newblock {\em Phys. Rev. Lett}, vol. 69(20), pp. 2881--2884, 1992.

\bibitem{CDNT97}
R.~Cleve, W.~van Dam, M.~A. Nielsen, and A.~Tapp,
\newblock ``Quantum entanglement and the communication complexity of the inner
  product function,''
\newblock in {\em Proceedings of the 1st NASA International Conference on
  Quantum Computing and Quantum Communications, Lecture Notes in Computer
  Science}. 1998, vol. 1509, pp. 61--74, Springer-Verlag,
\newblock quant-ph/9708019.

\bibitem{FF56}
L.~R.~Ford{,}~Jr. and D.~R.~Fulkerson,
\newblock {\em Flows in networks},
\newblock Princeton University Press, 1962.

\bibitem{Gottesman99qss}
D.~Gottesman,
\newblock ``On the theory of quantum secret sharing,''
\newblock {\em Phys. Rev. A}, vol. 61, pp. 042311, 2000,
\newblock quant-ph/9910067.

\bibitem{IMNTW03}
H.~Imai, J.~Mueller-Quade, A.~C.~A. Nascimento, P.~Tuyls, and A.~Winter,
\newblock ``A quantum information theoretical model for quantum secret sharing
  schemes,''
\newblock {\em Quant.~Inf.~Comp.}, vol. 5(1), pp. 68--79, 2005.

\bibitem{supp1}
  Let $S(\cdot)$ denote the von Neumann entropy, and $I^{\rm coh}(S_1 \> S_2)
  = S(S_2) - S(S_1S_2)$ be the coherent information from $S_1$ to $S_2$. Let
  $S$ be the secret, purified by the reference system $R$, encoded into a
  secret sharing scheme, and to be decoded in $\tilde{S}$. If the initial state
  in $SR$ and the final state in $\tilde{S}R$ are $\epsilon$-close to each
  other in trace distance, then the coherent information $I^{\rm coh}(R \>
  \tilde{S})$ has to be $\epsilon'$-close to $I^{\rm coh}(R \> S)$. (Note that
  $\epsilon'$ and $\epsilon$ are related by Fannes inequality. For constant
  size secret, they vanish together. For asymptotically large secret of
  dimension $2^{rn}$, where $n$ is the number of uses of the network and $r$ is
  the rate of transmission, $\epsilon' \approx n \epsilon$ with effects on the
  communication rate being $\approx \epsilon'/n \approx \epsilon \rightarrow
  0$. That $\epsilon' \rightarrow 0$ is often used as an alternative condition
  for preserving entanglement fidelity.) Let $S_s$ be the significant share,
  and $S_u$ be the unauthorized set such that $S_s \bigcup S_u$ is authorized.
  In other words, $\tilde{S}$ is obtained from $S_s S_u$ by a quantum
  operation. But the coherent information is monotonically decreasing under
  such operations. Therefore, \be I^{\rm coh}(R \> S_s S_u) \geq I^{\rm coh}(R
  \> \tilde{S}) \geq I^{\rm coh}(R \> S) - \epsilon' \,. \ee Using the
  expression of $I^{\rm coh}$ and the joint purity of $RS$, \begin{equation}
  S(S_s S_u) - S(R S_s S_u) \geq S(S) - \epsilon' \,. \end{equation} Applying
  to the above the Araki-Lieb inequality, which states that $\forall_{S_1 S_2}
  S(S_1S_2) \geq |S(S_1)-S(S_2)|$, with $RS_u = S_1, S_s = S_2$, we have
  \begin{equation} S(S_s S_u) - [S(R S_u) - S(S_s)] \geq S(S) - \epsilon'
  \end{equation} Since $S_u$ is unauthorized, the mutual information between
  $S_u$ and $R$ is small, so we have $S(S_u) + S(R) - S(R S_u) \leq \gamma$ for
  some small $\gamma$. Substituting this in the above, \begin{equation} S(S_s
  S_u) + \gamma - [S(S_u) + S(R)] + S(S_s) \geq S(S) - \epsilon' \end{equation}
  Finally, applying subadditivity $S(S_s S_u) \leq S(S_u) + S(S_s)$ to the LHS,
  and noting that $S(R) = S(S)$, and rearranging terms, \begin{equation} S(S_s)
  \geq S(S) - (\epsilon' + \gamma)/2 \end{equation}

\bibitem{BHJW00}
H.~Barnum, R.~Josza, P.~Hayden, and A.~Winter,
\newblock ``On the reversible extraction of classical information from a
  quantum source,''
\newblock {\em Proc.~Roy.~Soc.~(Lond.) A}, vol. 457, pp. 2019--2039, 2001.

\bibitem{supp2}
  All notations are as defined in the statement of the result in the main
  text. Let $\stackrel{\epsilon}{\approx}$ denote the relation between two
  states differing by at most $\epsilon$ in trace distance. The isometry $Y$
  can be represented as a unitary taking $AC$ to $\tilde{S}E$, with $C$
  initially in some fixed state $|0\>$. Then, the approximate invertibility
  condition can be stated as: \begin{eqnarray} \forall \, |\psi\>_{RS} & &
  (I_{RBD} \otimes Y_{AC}) (I_R \otimes W_S) |\psi\>_{RS} \, |0\>_C \nonumber
  \\ & \stackrel{\epsilon}{\approx} & |\psi\>_{R\tilde{S}} \, |\phi\>_{BDE}
  \label{eq:invert} \end{eqnarray} Equation (\ref{eq:invert}) holds when the
  systems $BDE$ are traced out, due to the notion of approximation in the
  achievable rate region. Without tracing out $BDE$, the global state is pure,
  and Eq.~(\ref{eq:invert}) follows from Ulhmann's theorem. Note that the same
  state $|\phi\>_{BDE}$ appears on the RHS for all $|\psi\>_{RS}$, or else
  $|\phi\>_{BDE}$ allows extraction of more information about the state
  $|\psi\>_{RS}$ beyond what is allowed by the error $\epsilon$. We will use
  Eq.~(\ref{eq:invert}) in two different ways. First, due to the universal
  quantifier, we apply Eq.~(\ref{eq:invert}) to the state $(I_R \otimes U_S)
  |\psi\>_{RS}$ instead. \begin{eqnarray} \forall \, |\psi\>_{RS} && (I_{RBD}
  \otimes Y_{AC}) (I_R \otimes W_S U_S) |\psi\>_{RS} |0\>_C \nonumber \\ &
  \stackrel{\epsilon}{\approx} & (I_{RBDE} \otimes U_{\tilde{S}}) \, [
  |\psi\>_{R\tilde{S}} \, |\phi\>_{BDE} ] \label{eq:invert1} \end{eqnarray}
  Second, we replace the expression in the square bracket above by using
  Eq.~(\ref{eq:invert}): \begin{eqnarray} \forall \, |\psi\>_{RS} \hspace*{3ex}
  (I_{RBD} \otimes Y_{AC}) (I_{R} \otimes W_S U_S) |\psi\>_{RS} |0\>_C
  \label{eq:invert2} \end{eqnarray} \vspace*{-6ex} \begin{eqnarray}
  \hspace*{10ex} \stackrel{2\epsilon}{\approx} (I_{RBDE} \otimes 
  U_{\tilde{S}}) (I_{RBD}
  \otimes Y_{AC}) (I_R \otimes W_S) |\psi\>_{RS} |0\>_C \nonumber
  \end{eqnarray} where the error above is obtained by using the triangle
  inequality for the trace distance, and it is at most the sum of the errors in
  Eqs.~(\ref{eq:invert}) and (\ref{eq:invert1}). Now, applying $Y^\dagger$ 
  to both
  sides of Eq.~(\ref{eq:invert2}), we obtain \begin{eqnarray} \forall \,
  |\psi\>_{RS} ~~~~ (I_R \otimes W_S U_S) |\psi\>_{RS} |0\>_C \hspace*{15ex}
  \nonumber \end{eqnarray} \vspace*{-6ex} \begin{eqnarray} \nonumber
  \hspace*{15ex} \stackrel{2\epsilon}{\approx} (I_{RBD} \otimes Y_{AC}^\dagger)
  (I_{RBDE} \otimes U_S) (I_{RBD} \otimes Y_{AC}) \\ \nonumber
  \hspace*{35ex}\times (I_R \otimes W_S) |\psi\>_{RS} |0\>_C \end{eqnarray}
  which is the result we asserted.

\bibitem{Holevo73}
A.~S. Holevo,
\newblock ``Bounds for the quantity of information transmitted by a quantum
  communication channel,''
\newblock {\em Problemy Pere\-dachi Informatsii}, vol. 9(3), pp. 3--11, 1973,
\newblock [A. S. Kholevo, {\it Problems of Information Transmission\/}, vol.\
  9, pp.\ 177-183 (1973)].

\bibitem{H07}
M.~Hayashi,
\newblock ``Prior entanglement between senders enables perfect quantum network
  coding with modification,''
\newblock {\em Phys. Rev. A}, vol. 76, 2007,
\newblock arXiv/0706.0197.

\bibitem{SVW05}
J.~A. Smolin, F.~Verstraete, and A.~Winter,
\newblock ``Entanglement of assistance and multipartite state distillation,''
\newblock {\em Phys. Rev. A}, vol. 72, pp. 052317, 2005,
\newblock quant-ph/0505038.

\bibitem{HOW05s}
M.~Horodecki, J.~Oppenheim, and A.~Winter,
\newblock ``Partial quantum information,''
\newblock quant-ph/0505062, 2005.

\bibitem{HOW05l}
M.~Horodecki, J.~Oppenheim, and A.~Winter,
\newblock ``Quantum state merging and negative information,'' quant-ph/0512247.

\bibitem{FL06}
B.~Fortescue and H.-K. Lo,
\newblock ``Random bipartite entanglement from $w$ and $w$-like states,''
\newblock quant-ph/0607126, 2006.

\bibitem{DFMSTU98}
D.~P. DiVincenzo, C.~A. Fuchs, H.~Mabuchi, J.~A. Smolin, A.~Thapliyal, and
  A.~Uhlmann,
\newblock ``Entanglement of assistance,''
\newblock in {\em Proceedings of the 1st NASA International Conference on
  Quantum Computing and Quantum Communications, Lecture Notes in Computer
  Science}. 1998, vol. 1509, Springer-Verlag,
\newblock quant-ph/9803033.

\bibitem{PVMC04}
M.~Popp, F.~Verstraete, M.~A. Martin-Delgado, and J.~I. Cirac,
\newblock ``Localizable entanglement,''
\newblock {\em Phys. Rev. A}, vol. 71, pp. 042306, 2005,
\newblock quant-ph/0411123.

\end{thebibliography}

\end{document}